\documentclass[11pt]{article}
\usepackage{graphicx}
\usepackage{wrapfig}
\begin{document}

\begin{center}
\vspace*{0.2in}\begin{LARGE}\begin{bf}
Interplay between magnetic and spatial\\[1mm] order in quasicrystals
\end{bf}
\end{LARGE}
\vspace{8mm}

{\large E.Y.~VEDMEDENKO$^{\star}$,
U.~GRIMM$^{\dag}$\footnote{Corresponding author. Email:
u.g.grimm@open.ac.uk} and R.~WIESENDANGER$^{\star}$}\vspace{3mm}

$\mbox{}^{\star}$Institut f\"{u}r Angewandte Physik,
Universit\"{a}t Hamburg,\\
Jungiusstr. 11, 20355 Hamburg, Germany\\
$\mbox{}^{\dag}$Applied Mathematics Department,
The Open University,\\ Walton Hall, Milton Keynes MK7 6AA, UK
\end{center}\vspace{8mm}

\begin{small}\parbox{5in}{ The stable
magnetisation configurations of antiferromagnets on quasiperiodic
tilings are investigated theoretically. The exchange coupling is
assumed to decrease exponentially with the distance between magnetic
moments. It is demonstrated that the combination of geometric
frustration and the quasiperiodic order of atoms leads to
complicated noncollinear ground states. The structure can be divided
into subtilings of different energies. The symmetry of the
subtilings depends on the quasiperiodic order of magnetic moments.
The subtilings are spatially ordered. However, the magnetic ordering
of the subtilings in general does not correspond to their spatial
arrangements. While subtilings of low energy are magnetically
ordered, those of high energy can be completely disordered due to
local magnetic frustration. \vspace{8mm}

{\it Key words:} Quasicrystals; Aperiodic tilings; Magnetic order;
Frustration}\end{small}
\vspace{1cm}

\subsection{Introduction}

\hspace*{\parindent}
In contrast to the rather well-studied spin structure of
antiferromagnets on periodic lattices, the antiferromagnetic ordering
of quasicrystals is subject of ongoing scientific debate
\cite{LTC93,L95,I98,H00,SS00,STTSOA00,DJSGT03,WJH03,VOK03,VGW04,J04}.
Experimentally, it has been demonstrated that rare earth containing
quasicrystals exhibit spin glass-like freezing at low temperatures
\cite{I98,SS00}. However, this freezing is different from that of
conventional spin glasses. The observed dependence of the
thermoremanent magnetisation on the magnetic field does not follow the
spin-glass behaviour and the frequency shift of the freezing
temperature lies between those of a canonical spin glass and of a
superparamagnet \cite{DJSGT03}. Hence, the free energy landscape of a
rare earth quasicrystal is different from both the highly degenerate
distribution of energy barriers in spin glasses and the single global
energy minimum in superparamagnets.

\begin{figure}[t]
\centerline{\includegraphics{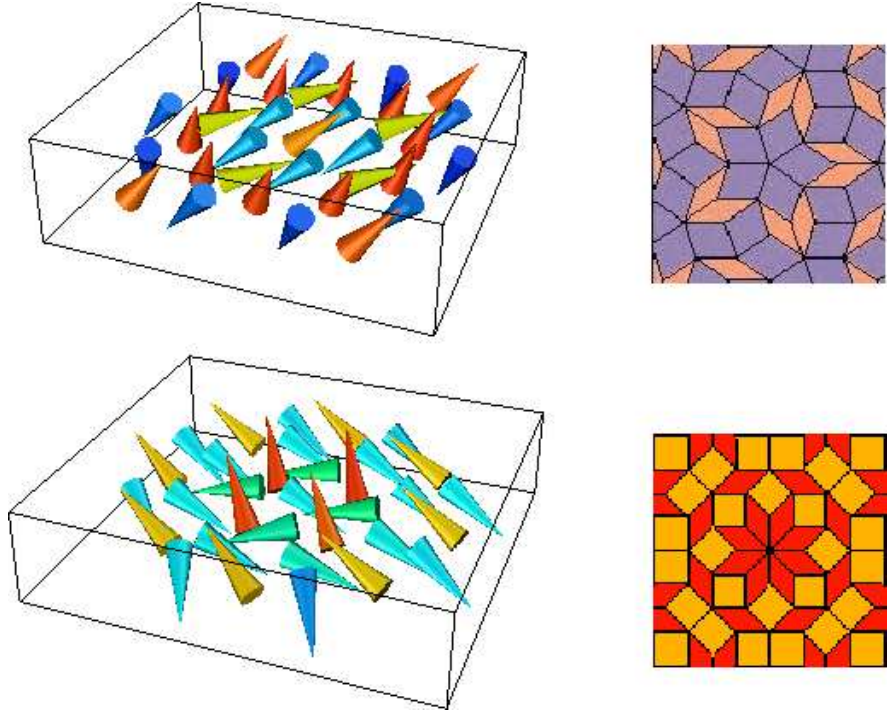}}
{\caption{Perspective view of a portion of a Monte-Carlo
configuration on the Penrose tiling (top) and the octagonal
tiling (bottom). Top views of the corresponding patches are
shown on the right. The magnetic moments are
represented as cones.} \label{Abb1}}
\end{figure}

Although the atomic and electronic structure of rare earth
quasicrystals is not completely understood, it has been postulated
\cite{DJSGT03} that the low-temperature microstructure of such a
magnet resembles geometrically frustrated but site-ordered magnetic
systems and consists of weakly interacting magnetically ordered
clusters. Another interesting approach is based on recent elastic
neutron scattering experiments on a Zn-Mg-Ho icosahedral
quasicrystal \cite{STTSOA00} revealing a very peculiar diffuse
scattering pattern with icosahedral symmetry at temperatures below
6K. In contrast to reference \cite{DJSGT03}, the authors interpret
the diffraction pattern as that of several interpenetrating
quasiperiodic sublattices, where all spins point in the same
direction \cite{VGW04}.  Recent theoretical studies of real-space
magnetic configurations on the octagonal tiling
\cite{WJH03,VGW04,J04} demonstrate that the energy landscape, in
accordance with \cite{DJSGT03}, is neither degenerate nor has a
single global minimum. All spins can be divided into several
quasiperiodic (in the 2D physical space) or periodic (in the
corresponding 4D periodic hypercrystal) subtilings of different
energy.

\begin{figure}[t]
\centerline{\includegraphics{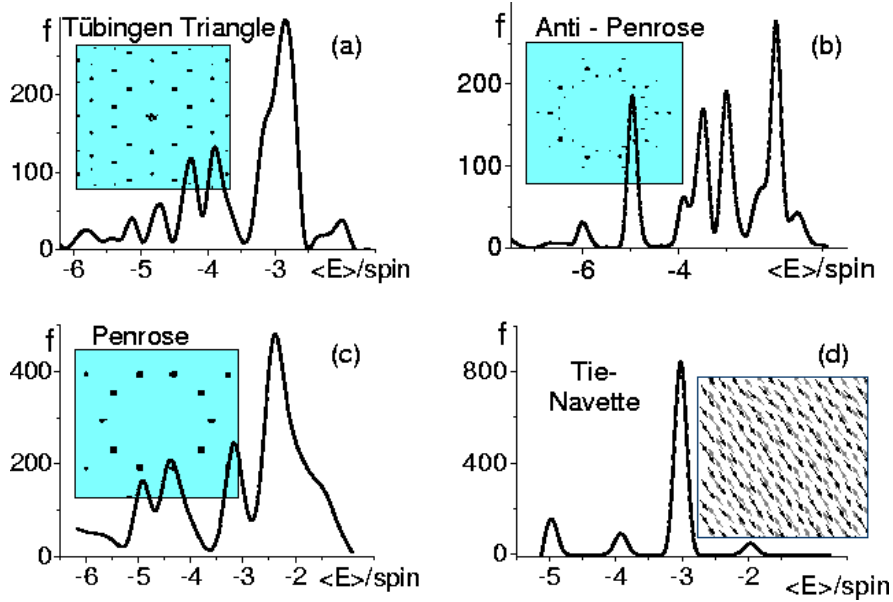}}
{\caption{The frequency distribution of the energy per spin on the
T\"{u}bingen triangle (a), Anti-Penrose (b), Penrose (c) and Tie-Navette
(d) tilings for classical vector spins. A purely antiferromagnetic
interaction $J$ at a temperature $kT=0.01J$ is considered.
The insets in (a)--(c) give the calculated Bragg scattering of
the $S^y$ component of the magnetisation for subtilings composed of magnetic
moments belonging to peaks with $-6<\frac{{\langle E\rangle
}}{{spin}}<-4$. The scale goes from -6 to 6 $k^{S_y}_{x,y}/\pi$. The
inset in (d) shows a portion of the stable magnetic configuration on
the Tie-Navette tiling as described in the text. Dark and light grey
arrows denote antiparallel magnetic moments.} \label{Abb2}}
\end{figure}

In the present investigation, we calculate the low-temperature stable
antiferromagnetic configurations on several planar quasiperiodic
tilings with tenfold symmetry. In most rare earth intermetallic
compounds an oscillatory (RKKY - like) exchange interaction has been
observed. To tackle this complicated problem first we concentrate on
exponentially decreasing exchange coupling corresponding to a
rapid-decaying limit of an oscillatory interaction. It will be
demonstrated that the real-space magnetic structure is generally
three-dimensional and noncollinear. In disagreement with
\cite{DJSGT03}, and in accordance with \cite{STTSOA00}, the magnetic
structure consists of several ordered interpenetrating quasilattices
with characteristic wave vectors.

\subsection{Simulations and results}

We have investigated the magnetic ordering in an antiferromagnet on
Penrose, Anti-Penrose, T\"{u}bingen triangle \cite{BKSZ90} and
Tie-Navette \cite{LL94} tilings by means of Monte-Carlo
simulations. Two-dimensional films of classical, three-dimensional
magnetic moments $\mathbf{S}$ have been studied.  The Hamiltonian of
the problem is given by
\begin{equation}\label{1}
H = J_{ij}\sum\limits_{\langle i,j\rangle}
\mathbf{S}_{i} \cdot \mathbf{S}_{j} - K_{1}\sum\limits_{i}
(\mathbf{S}_{i}^{z})^2
\end{equation}
where $J_{ij}$ are the exchange coupling constants and $\langle
i,j\rangle$ refers to pairs of spins.  Two cases have been explored:
$J_{ij}=1$ for all $r_{ij}\le 1$ (and $J_{ij}=0$ for all $r_{ij}>1$),
and an exponential decrease of the exchange interaction with the
distance between magnetic moments (which for practical purposes was
cut off at distance $r_{ij}>2$), where $r_{ij}$ denotes the distance
between sites $i$ and $j$ (as compared to the edge length in the
tiling, which are chosen to have length one). The samples are patches
of square or rectangular shape, containing some $10\,500$ magnetic
moments. We also used circular areas to check that our results are not
affected by the shape of the sample. An extremely slow annealing
procedure, with $50$ temperature steps per Monte-Carlo run, has been
applied. To see the time-dependent changes in a microstructure, we ran
the simulation for several hundred thousand steps per temperature.

In previous theoretical studies \cite{LTC93,L95,H00} frustrated,
two-dimensional structures have been proposed. In accordance with
previous publications, we find that the ground state of a system
with purely antiferromagnetic exchange interactions is
locally frustrated. Under the local frustration $f$ we
understand the normalised difference between an actual energy $E_i$
of a spin $i$ and a ground state energy $E_{id}$ of a relevant
unfrustrated vertex with all spins antiparallel to the spin $i$
\begin{equation}\label{2}
f = \frac{{|E_{id} | - |E_i |}}{{|E_{id} |}}.
\end{equation}
In contrast to the common folklore, the configurations are
three-dimensional. Similar to the underlying atomic symmetry, the
magnetic structure is quasiperiodic, i.e.\ it consists of identical
units which do not have identical surroundings.

\begin{figure}[t]
\centerline{\includegraphics{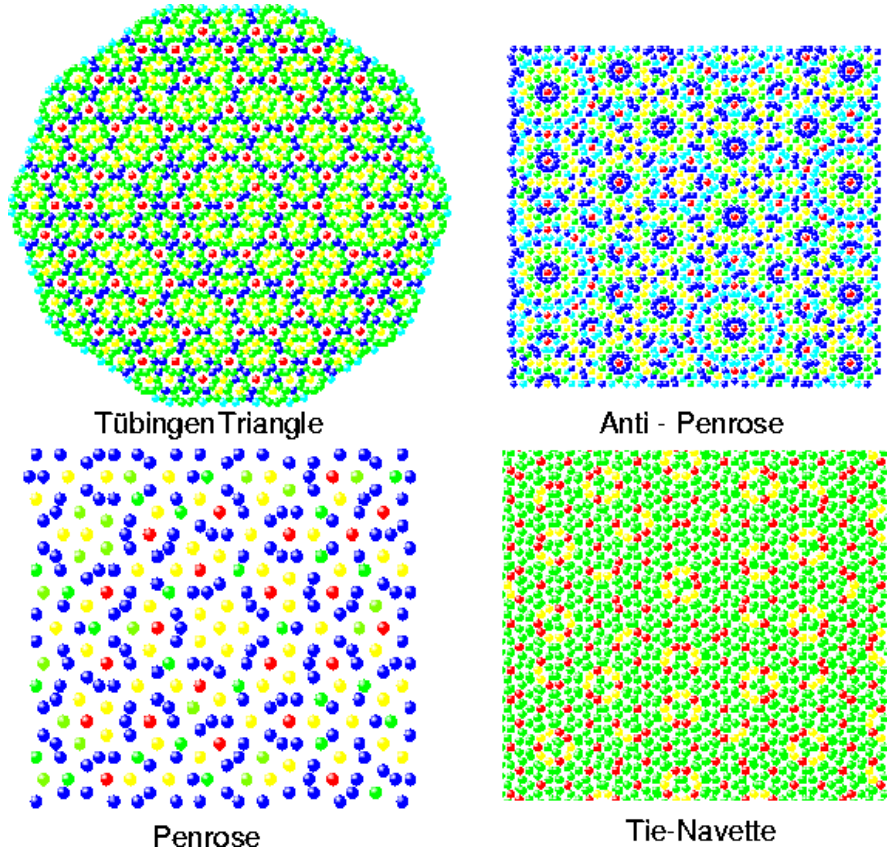}}
{\caption{Energy maps for classical vector spins on T\"{u}bingen
triangle (a), Anti-Penrose (b), Penrose (c) and Tie-Navette (d)
tilings. The circles give positions of magnetic moments. Different
shades of grey denote different energies corresponding to the peaks in
figure~\ref{Abb2}. Purely antiferromagnetic interaction with $J=1$ for
all $r_{ij} \leq1$ at $kT=0.01J$ is considered.}
\label{Abb3}}
\end{figure}

Three-dimensional representations of parts of the low-temperature
quasiperiodic patterns observed for the Penrose and the octagonal
tiling are shown in figure~\ref{Abb1}. The corresponding
configurations represent the characteristic Penrose and Amman-Beenker
`stars', which are also shown in figure~\ref{Abb1} for clarity. On the
Penrose tiling, the `star'-pattern can easily be recognised in the
magnetic structure, because the moments belonging to the perimeter of
enclosed `stars' show perfectly antiparallel alignment.  On the
octagonal tiling, the situation is more complicated. The central
magnetic moment is neither parallel nor antiparallel to the
neighbouring magnetic moments. Its eight nearest neighbours have
different sets of mutual angles. The moments forming the next ring
have still another orientation with respect to their nearest
neighbours. The noncollinear alignment of the neighbouring moments
indicates that the system is geometrically frustrated, i.e.\ there is
no possibility to align all neighbours in an antiparallel
arrangement. Similar noncollinear antiferromagnetic configurations are
formed in the T\"{u}bingen triangle and Anti-Penrose tilings. Within
the examples of tilings considered here, the Tie-Navette tiling
represents an exception. The magnetic structure observed for this
tiling consists of two antiferromagnetically aligned quasiperiodic
sublattices, as shown in figure~\ref{Abb2}(d). This means that every
pair of nearest neighbouring moments can be aligned antiparallelly,
i.e.\ the antiferromagnetic configuration is not frustrated.

We have calculated the stable low-temperature configurations and the
frequency distribution of the exchange energy per atom $\langle E
\rangle$ for the T\"{u}bingen triangle, Anti-Penrose, Penrose and
Tie-Navette tilings. The calculations have been performed for an
exponentially decreasing exchange coupling and for a short-range
exchange coupling $J_{ij}=\mbox{const}=1$ for all $r_{ij}\leq 1$.
The analysis of the local energies reveals several characteristic
energetic maxima in the frequency distributions shown in
figure~\ref{Abb2}(a)--(d). The magnetic configurations and the
number of the energy peaks for the same tiling are identical for
both choices of exchange couplings ($J_{ij}\propto e^{-r_{ij}}$ and
$J_{ij}=1$ for $r_{ij}\le 1$). For different tilings, the number and
the width of the maxima are different. The simple existence of the
peaks means that there exist different sorts of magnetic moments
having well-defined relative orientations with respect to their
nearest neighbours. These relative orientations depend on the tiling
and {\em not}\/ on the choice of the exchange couplings $J_{ij}$.
For $J(r_{ij}\le 1)=1$, however, it can be seen directly from the
energy distributions of figure~\ref{Abb2}, whether the magnetic
ordering is collinear or noncollinear. If all nearest neighbours are
collinear (parallel or antiparallel), then the exchange energy per
spin should have integral values depending only on the number of the
neighbouring moments. This is indeed the case for the Tie-Navette
tiling; compare figure~\ref{Abb2}(d). For a noncollinear alignment
of neighbouring magnetic moments, $\left\langle E \right\rangle $
should be non-integral as the cosines of the angles between the
moments are no longer zero or unity. This happens for all other
tilings we considered; compare figure~\ref{Abb2}(a)--(c). The
average energy of noncollinear configurations is smaller than the
energy of any collinear solution.  Hence, the increase of the
configurational entropy permits to minimise the average
local frustration and the total energy of the system.

The spatial arrangements of the exchange energies of the magnetic
moments are given in figure~\ref{Abb3}. Each shade of grey in
figure~\ref{Abb3} represents a certain energy range corresponding to
one of the peaks in the spectra of figure~\ref{Abb2}. The magnetic
moments form subtilings of different energies, which generally do not
coincide with a tiling obtained by selecting a specific vertex
type. The subtilings of low energy $\frac{{\langle
E\rangle }}{{spin}}<-3$ are magnetically stable and ordered while
those of higher energy $\frac{{\langle E\rangle }}{{spin}}>-3$
disordered. The disorder can be seen in the portion of the magnetic
configuration shown at the bottom of figure~\ref{Abb1}. The two front
moments belonging to the subtiling of a large energy have angles which
deviate considerably from those of the other moments in the ring while
the moments in the inner rings with lower energy have collinear
orientations. With increasing temperature the magnetisation of
subtilings of large energy is fluctuating while the magnetisation of
low-energy subtilings is still stable. The spatial quasiperiodic
ten-fold symmetry of the ordered subtilings can be seen from the
calculated magnetic Bragg scattering given in the insets to 
figure~\ref{Abb2}. While the atomic ordering of the unstable
subtilings can be seen in the Fourier space their magnetic reflexes
are extinct because of disorder.

\subsection{Summary}

In conclusion, we demonstrate that vector spin system with
antiferromagnetic coupling on different quasiperiodic tilings is
locally frustrated. All spins can be divided into several
quasiperiodic (in our two-dimensional physical space) or periodic (in
the corresponding four-dimensional periodic hypercrystal) subtilings
of different energy, which generally do not coincide with a specific
vertex type. The vector spin system admits a three-dimensional
noncollinear magnetic structure. The noncollinearity of the magnetic
configuration permits to minimise the degree of frustration and the
total energy of the system in comparison with the collinear case. The
co-directional spins of every subtiling reveal quasiperiodic ordering
with a wave vector which is specific for a given subtiling. The
Tie-Navette tiling is not frustrated and admits collinear magnetic
configurations. For the short-ranged exchange interaction, this arises
as a consequence of the bipartiteness of the graph formed by
connecting interacting pairs of spins; however, we observe that the
antiferromagnetic order persists for the case of a long-range,
exponentially decreasing exchange interaction.

\end{document}